\documentclass[12pt,preprint]{aastex}

\input epsf

\newcommand{\be}{\begin{equation}}
\newcommand{\ee}{\end{equation}}
\newcommand{\bea}{\begin{eqnarray}}
\newcommand{\eea}{\end{eqnarray}}

\received{6 May 1987}
\accepted{1 July 1988}
\journalid{}{}
\articleid{}{}

\lefthead{M.P. D\c{a}browski and J. Stelmach}
\righthead{Observable Quantities in Models with Strings}

\begin{document}

\title{Observable Quantities in Cosmological Models with Strings}

\author{Mariusz P.~D\c{a}browski and Jerzy Stelmach}

\affil{Institute of Physics, University of Szczecin, Wielkopolska 15,
70-451 Szczecin, Poland}

\begin{abstract}

The Friedman equation for the universe with arbitrary curvature $(k = 0, \pm 1)$,
filled with mutually noninteracting pressureless dust, radiation, cosmological constant,
and strings is considered. We assume the string domination
scenario for the evolution of the latter component. Moreover, we
discuss the simplest possibility for the scaling of the string
energy density: $\varrho \propto R^{-2}$. For such models we write
down the explicit solution of the Friedman equation. We realize
that corresponding cosmological models do not essentially differ
from those without strings. We find an analytic formula for the
radial coordinate $\chi$ of a galaxy with a redshift z and express
it in terms of astronomical parameters. This relation is then used
for the derivation of the astrophysical formulas such as
luminosity distance, angular diameter, and source counts, which
may serve for testing the string-dominated universe. It seems that
the most sensitive test, at least from the formal point of view,
is the formula for the number of galaxies $N(z)$ corresponding to
a given value of the redshift. We show that the maximum of $N(z)$
strongly depends on the density of strings, especially if the
density is large enough to explain the $\Omega$ problem. Other
tests are also proposed.

\end{abstract}


{\bf Journal Reference: The Astronomical Journal {\bf 97}, 978 (1989).}

\section{Introduction}

One of the consequences of the inflationary scenario is the near
flatness of the universe, which means that the energy density is
very close to the critical value $\Omega = \varrho/\varrho_c = 1$
(Guth 1981; for observational aspects, see also Loh 1986; Loh and
Spillar 1986a,b). However, observation of galaxies gives us a
value roughly one order less. A lack of observable mass necessary
for approaching the critical density is usually called the dark
matter problem. people propose many solutions to this problem
(Turner 1987). One of the simplest solutons is the assumption that
the cosmological constant does not vanish. Other possibilities are
massive neutrinos, axions, or heavy superpartners of the usual
particles, for example, photinos. Several years ago, when the
theories of grand unification (GUTs) were used for the description
of the early universe, one more candidate appeared to solve the
mentioned problem - cosmic strings.

In gauge theories with spontaneous symmetry breaking, the phase
transition at critical temperature (above which symmetry can be
restored) can give rise to the nontrivial vacuum structure of the
universe (Zel'dovich {\it et al.} 1974). These are domain walls,
strings, or monopoles, depending on the topology of the manifold
of degenerate vacua (Kibble 1976). More complicated topological
objects such as walls bounded by strings or monopoles connected by
strings can also be formed (Vachaspati and Vilenkin 1984). However,
we have good reason to believe that, up to the present time, in
our visible part of the universe only strings, or eventually an
infinite network of strings with monopoles in vertices, could
survive (Vilenkin 1981a; Vachaspati and Vilenkin 1987).

It is well-known that there are two extreme configurations to
which a system of strings may evolve: a scaling configuration or a
string-dominated universe (SDU). The former possibility has been
extensively examined in the context of galaxy formation
(Zel'dovich 1980; Vilenkin 1981b; Vilenkin and Shafi 1983; Turok and
Brandenberger 1986; Kibble 1986). The latter one is also very
attractive because it may give the explanation for the missing
mass (Vilenkin 1984a; Turner 1985; Kibble 1986). We shall not
discuss both scenarios, referring instead to the literature
(Vilenkin 1985; Kibble 1985; Bennet 1986a,b; Scherer and Frieman
1986; Aryal {\it et al.} 1986). Our considerations are based on
the result of Turok and Bhattacharjee (1984), who have shown that,
neglecting interaction, the energy density of a network of strings
scales as $\varrho_s \propto R^{-n}$, where $2\leq n \leq 3$ (see
also Kibble (1986) and Vachaspati and Vilenkin (1987). The case $n=2$
and the astronomical constraints on SDU have been examined by Gott
and Rees (1987). Earlier, Gott (1985) had given exact solution and
presented interesting considerations concerning the gravitational
field of strings in the context of gravitational lensing (see also
Hiscock (1985)).

The purpose of the present paper is to calculate some atrophysical
formulas for the homogeneous and isotropic universe with arbitrary
curvature $(k=0,\pm 1)$ filed with pressureless dust, radiation,
cosmological constant, and the system of strings. For simplicity,
we consider only an extreme case when $n=2$ (it corresponds, for
example, to the set of randomly oriented straight strings or to
the tangled network of strings which conformally stretches by the
expansion). This case is particularly interesting because it
allows treatment of all mentioned components of the universe
simultaneously in an analytic way. Generally, this is not the case
if $2\leq n \leq 3$. Although the case under consideration is
probably too idealistic, it gives some aspects of observational
problems in the universe with strings and in some sense completes
the discussion of the string-dominated universe given by Vilenkin,
Kibble, and Kardashev (1986), Charlton and Turner (1987), or Gott
and Rees (1987).

Ther exist several observational tests of the cosmic-string
scenario: gravitational lensing (Vilenkin 1984b), anisotropy in
the microwave-background radiation (Kaiser and Stebbins 1984), and
gravitational radiation emitted by decaying loops (Hogan and Rees
1984). The tests proposed in the present paper are of a different
type and are not relevant for detecting a single string. Our tests
could be applied only for the universe in which the network of
strings is sufficiently dense in order to influence astrophysical
observables such as luminosity distance, angular diameter, or
source counts. Moreover, the strings under consideration should be
relatively light (Vilenkin 1984a).

We do not want to make a statement pro or con regarding the string
scenario. Our task is to propose tests that could be used for the
verification of the hypothesis concerning a universe dominated by
a network of strings. It seems to us that at least one of them,
namely source counts, is very promising for this purpose. We leave
consideration of the less trivial case of the string-domination
scenario (where $3>n>2$ and the universe behaves very like a matter-dominated
one) for the future.

In the next section, we introduce some definitions and we describe
the Friedmann equation for the universe with strings in a form
that allows use of the methods and results (with some
modifications) from our previous papers concerning the case
without strings (D\c{a}browski and Stelmach 1986a,b; 1987a). Next,
we give analytic solutions of the Friedman equation in terms of
Weierstrass elliptic and nonelliptic functions. In Sec. III, we
find the relations between astronomical parameters and derive some
astrophysical formulas for the universe with strings. Section IV
is devoted to the discussion of the tests for the existence of the
network of cosmic strings. We pay much attention to the case where
the formulas are given by elementary functions. Because of their
simplicity, these cases are of special interest. In Sec. V we
summarize the results.

\section{Friedman Equation Including Strings}

We consider Friedman models described by the equation
\be
\label{FRW1}
\dot{R}^2 + k = \frac{8\pi G}{3} \varrho R^2~,
\ee
where $\varrho$ is total energy density of the universe including,
besides the usual components $\varrho_m$ -- matter, $\varrho$ --
radiation, $\varrho_{\Lambda}$ -- cosmological constant), also
strings $\varrho_s$:
\be
\varrho = \varrho_r + \varrho_m + \varrho_s + \varrho_{\Lambda}~.
\ee
$k=0,\pm1$ is the curvature index and $R(t)$ is the scale factor.
We assume that strings satisfy the equation of state (Zel'dovich
1980, Vilenkin 1981c)
\be
p_s = -\frac{\varrho_s}{3}~.
\ee
Noninteraction between the components gives the simple relation
for $\varrho_s$, namely
\be
\label{Cs}
\varrho_s R^2  =  C_s \frac{3}{8\pi G}~.
\ee
Following the notation of Coquereaux and Grossmann (1982), we give
the similar expressions for $\varrho_r,\varrho_m$, and
$\varrho_{\Lambda}$:
\bea
\label{CCC}
\varrho_r R^4 & = & C_r \frac{3}{8\pi G}~  ,\\
\varrho_m R^3 & = & C_m \frac{3}{8\pi G}~  ,\\
\varrho_{\Lambda} & = & \frac{3}{8\pi G}~  ,
\eea
where $C_r$, $C_m$, and $C_s$ are constants.
If we now introduce a new parameter
\be
\label{k'}
k' \equiv k - C_s~
\ee
Friedmann equation (\ref{FRW1}) acquires a form identical to that
without strings with the only replacement $k \to k'$
\be
\label{FRW2}
\dot{R}^2 + k = \frac{C_m}{R} + \frac{C_r}{R^2} + \frac{\Lambda}{3}
R^2~.
\ee
However, generally the sign of $k'$ now has nothing to do with the
curvature of the universe.

The formal resemblance of both models (with and without strings)
notably simplifies treatment of the present case. In deriving
different expressions for the models without strings
(D\c{a}browski and Stelmach 1986a,b; 1987b), for the most part we
did not employ the exact value of $k$. Thus, almost all formulas
are still valid if we put $k'$ instead of $k$. For example, Eq.
(\ref{FRW2}) rewritten in terms of dimensionless variables and
parameters takes the form
\begin{equation}
\label{FRWT}
\left(\frac{dT}{d\tau}\right)^2=\alpha T^4+\frac{2}{3}T^3-k' T^2+
\frac{\lambda}{3},
\end{equation}
where
\bea
\label{ceemetc}
\lambda &=& \frac{\Lambda}{\Lambda_{c}}~,\\
\Lambda_c &=& \frac{4}{(9C_{m}^2)}~,\\
\alpha &=& C_{r}\Lambda_{c}~,\\
d\tau &=& \frac{dt}{R}~,\\
T(\tau) &=& \Lambda_c^{-\frac{1}{2}}R^{-1}(\tau).
\eea

The solution of the Friedman equation may now be written down in a
parametric representation in at least two ways. In the first
version, we employ the Weierstrass elliptic function ${\cal P}$,
given as the solution of a differential equation (Tricomi 1937;
Abramovitz and Stegun 1964)
\be
\label{Weierstrass}
\left( \frac{d{\cal P}}{d\tau} \right)^2 = 4{\cal P}^3 -
g_{2}{\cal P} - g_{3}~   ,
\ee
where
\be
\label{g2}
g_{2}  =  \frac{k'^2}{12} + \frac{\alpha\lambda}{3}~,
\ee
and
\be
\label{g3}
g_{3}  =  6^{-3} \left( k'^3 - 2\lambda \right) -
\frac{\alpha \lambda k'}{18}   .
\ee
Then, the solution reads
\begin{equation}
\label{Rtaua}
R(\tau) = \frac{1}{6\sqrt{\Lambda_c}} \frac{3\sqrt{\alpha} {\cal P}'(\tau) + {\cal P}(\tau)
+ \frac{k'}{12}}{\left[{\cal P}(\tau) + \frac{k'}{12} \right]^2 - \frac{\alpha\lambda}{12}}~.     ,
\end{equation}
In the second case, nonelliptic functions $\zeta$ and $\sigma$ are
used:
\begin{equation}
\label{Rzeta}
R(\tau) = \sqrt{\frac{3}{\Lambda}} \left[ \zeta(\tau - \tau_{g}) - \zeta(\tau + \tau_{f})
+ \zeta(\tau_{g}) - \zeta(\tau_{f}) \right]~,
\end{equation}
\begin{equation}
\label{tauzeta}
\left( \frac{\Lambda}{3} \right)^{\frac{1}{2}}
t(\tau) =
\tau \left[ \zeta(\tau_{g}) - \zeta(\tau_{f}) \right] +
\ln{\mid \frac{\sigma(\tau - \tau_g)\sigma(\tau_f)}{\sigma(\tau -
\tau_f)\sigma(\tau_g)}\mid}~.
\end{equation}
The numbers $\tau_f$ and $\tau_g$, which are generally complex,
are given by the formulas (D\c{a}browski and Stelmach 1987a)
\bea
\label{tauf}
\tau_f &=& {\cal P}^{-1} \left( - \frac{k'}{12} - \frac{1}{2}
\sqrt{\frac{\alpha\lambda}{3}} \right)~,\\
\label{taug}
\tau_f &=& {\cal P}^{-1} \left( - \frac{k'}{12} + \frac{1}{2}
\sqrt{\frac{\alpha\lambda}{3}} \right)~.
\eea
The discussion of the solution proceeds analogously as in the case
without strings (Coquereaux and Grossmann 1982; D\c{a}browski and
Stelmach 1986a). For details, we refer to D\c{a}browski and
Stelmach (1987b). We realize that, in general, the types of the
solutions do not essentially differ from the case without strings
(when $k'=k=0, \pm 1$). The fundamental formal difference is that
for the universe with strings $k'$ is not normalized and may take
arbitrary values from the interval $(-\infty,1>$. However, it
should be stresses that some connection between $k$ and $k'$
exists and follows from Eq. (\ref{k'}). Namely, for $k'>0$, there
is only one possibility, $k=1$, for $k'=0$ the universe may be
closed $(k=1)$ or flat $(k=0)$, and finally for $k'<0$ three
possibilities may occur $(k=0, \pm 1)$.

At the end of this section we discuss some solutions that take on
an especially simple form. These are the cases when the Friedman
equation is explicitly integrable, i.e., the Weierstrass functions
${\cal P}$, $\zeta$, and $\sigma$ degenerate to elementary ones.
We shall not find all such solutions, but, because of their
particular simplicity, models with a vanishing cosmological
constant $(\lambda=0)$ are of special interest. We get oscillating
models for $k'>0$:
\bea
\label{Rtau+}
R(\tau) &=& \frac{1}{3k'\sqrt{\Lambda_c}}
\left(1 - \cos{\sqrt{k'}\tau} + 3 \sqrt{\alpha k'} \sin{\sqrt{k'} \tau}
\right)~,\\
t(\tau) &=& \frac{1}{3k'\sqrt{\Lambda_c}} \left[ \tau -
\frac{1}{\sqrt{k'}} \sin{\sqrt{k'}\tau} - 3 \sqrt{\alpha} \left(\cos{\sqrt{k'} \tau}
- 1 \right) \right]~,
\eea
and monotonic ones for $k'<0$,
\bea
\label{Rtau-}
R(\tau) &=& \frac{1}{3k'\sqrt{\Lambda_c}} \left(1 -
\cosh{\sqrt{-k'}\tau} + 3 \sqrt{-\alpha k'} \sinh{\sqrt{-k'}
\tau}\right)~,\\
t(\tau) &=& \frac{1}{3k'\sqrt{\Lambda_c}} \left[\tau -
\frac{1}{\sqrt{-k'}} \sinh{\sqrt{-k'}\tau} - 3 \sqrt{\alpha} \left(\cosh{\sqrt{-k'} \tau}
- 1 \right) \right]~,
\eea
and for $k'=0$,
\bea
\label{Rtau0}
R(\tau) &=& \frac{1}{6\sqrt{\Lambda_c}} \tau \left(\tau + 6
\sqrt{\alpha} \right)~,\\
t(\tau) &=& \frac{1}{2\sqrt{\Lambda_c}} \tau^2 \left(\frac{1}{9} \tau +
\sqrt{\alpha} \right)~.
\eea
The last two formulas, which describe either the flat universe
without strings $(k=0, C_s=0)$ or the closed universe with strings
$(k=1=C_s)$, follow from Eqs. (\ref{Rtau+}) and (\ref{Rtau-}) by
taking the limit $k'\to 0$. These are the cases in which radiation
and matter are negligible $(C_r=C_m=0)$, which can happen in a
vacuum or in a string-dominated universe. Explicit integration of
Eq. (\ref{FRW2}) gives for different values of $k'$ and $\Lambda$
solutions that are qualitatively the same as in the stringless
cosmology:
\bea
\label{R1}
R(t) &=& \sqrt{3k'/\Lambda} \cosh{\sqrt{\Lambda/3}
t},\hspace{0.2cm} {\rm for} \hspace{0.2cm} k', \Lambda >0~,\\
\label{R2}
R(t) &=& \sqrt{-3k'/\Lambda} \sinh{\sqrt{\Lambda/3}
t},\hspace{0.2cm} {\rm for} \hspace{0.2cm} k'<0, \Lambda >0~,\\
\label{R3}
R(t) &=& \exp{\sqrt{\Lambda/3} t}, \hspace{0.2cm} {\rm for}
\hspace{0.2cm} k'=0, \Lambda >0~,\\
\label{Rtaud}
R(t) &=& \sqrt{3k'/\Lambda} \sinh{\sqrt{-\Lambda/3}
t},\hspace{0.2cm} {\rm for} \hspace{0.2cm} \Lambda, k' >0~,
\eea
and, finally,
\be
\label{R5}
R(t) = \sqrt{-k'} t, \hspace{0.2cm} {\rm for} \hspace{0.2cm} \Lambda=0, k'
<0~.
\ee
For simplicity, we put integration constants equal to zero. Note
that the last formula describes also the asymptotic behaviour of
the monotonic model given by eqs. (\ref{Rtau-}). Some numerical
calculation for the above models, especially for that given by Eq.
(\ref{Rtaud}), were performed by Kardashev (1986).

\section{Astrophysical Formulas}

Detailed discussion of the relations between observable quantities
in the usual Friedmann models has been performed in our recent
paper (D\c{a}browski and Stelmach 1987a). It has been pointed out
in the present work that the extension to the case with cosmic
strings may be easily achieved by formal replacement $k\to k'$.
Definitive solution of the problem consists therefore in
expressing $k'$ in terms of astronomical parameters
$q_0, \sigma_{r0}, \sigma_{m0}, \sigma_{s0}$. The last parameter
did not come out so far in our papers. Its appearance follows from
the existence of cosmic strings in the model and it is defined
\be
\sigma_{s0} = \frac{4\pi G \varrho_{s0}}{3H_0^2}~,
\ee
where $H_0$ and $\varrho_{s0}$ are present values of the Hubble
constant and the energy density of strings, respectively. We
calculate $k'$ using the definition
\be
k' = k - C_s = k - 2 \sigma_{s0} H_0^2 R_0^2
\ee
and next ruling out $H_0^2R_0^2$ from the relation
\be
\label{H0R0}
H_0^2R_0^2 = \frac{k'}{4\sigma_{r0} + 3\sigma_{m0} - q_0 - 1}~.
\ee
Finally, $k'$ reads
\be
\label{k'k}
k' = \left(4\sigma_{r0} + 3\sigma_{m0} - q_0 - 1\right) \frac{k}{4\sigma_{r0} +
3\sigma_{m0} + 2 \sigma_{s0} - q_0 - 1}~.
\ee
Remaining parameters determining the model expressed in terms of
$\sigma_{r0}, \sigma_{m0}, \sigma_{s0}, H_0$, and $q_0$ are
\bea
\label{Lamcob}
\Lambda_c &=& \frac{H_0^2}{9\sigma_{m0}^2}
\left(\frac{4\sigma_{r0} + 3\sigma_{m0} - q_0 - 1}{k'}
\right)^3~,\\
\label{Lamob}
\Lambda &=& 3H_0^2 \left( 2\sigma_{r0} + \sigma_{m0} - q_0
\right)~,\\
\label{lamob}
\lambda &=& 27 \sigma_{m0}^2 \left( 2\sigma_{r0} + \sigma_{m0} - q_0
\right) \left(\frac{k'}{4\sigma_{r0} + 3\sigma_{m0} - q_0 - 1}
\right)^3~,\\
\label{alphaob}
\alpha &=& \frac{2\sigma_{r0}}{9\sigma_{m0}^2}
\frac{4\sigma_{r0} + 3\sigma_{m0} - q_0 - 1}{k'}~.
\eea
For completeness, we define $\sigma_{r0}, \sigma_{m0}, H_0$, and
$q_0$:
\bea
\sigma_{r0} &=& \frac{4\pi G \varrho_{r0}}{3H_0^2}~,\\
\sigma_{m0} &=& \frac{4\pi G \varrho_{m0}}{3H_0^2}~,\\
H_0 &=& \frac{\dot{R}}{R}~,\\
q_0 &=& - \frac{\ddot{R}_0 R_0}{\dot{R}_0^2}~,
\eea
where a zero means that the magnitudes correspond to the present
value of the cosmic time $t_0$. We note that application of the
formula (\ref{k'k}) to the expression (\ref{Lamcob}),
(\ref{lamob}), and (\ref{alphaob}) removes the explicit dependence
on $k'$, however, at the cost of also removing the proper
parameter describing strings $\sigma_{s0}$. From the same
formula,together with Eq. (\ref{H0R0}), a very interesting
property of the models with strings may be deduced, namely,
changing of the curvature of the universe without altering its
dynamics (Gott and Rees 1987). Gott and Rees come to this conclusion
by investigating the local influence of the strings on the
geometry of the universe. In our approach, the system of strings
forms a continuous fluid satisfying the exotic equation of state
$p = - \frac{1}{3} \varrho$. In this sense, we examine the global
influence of strings on the evolution. In order to see how the
above property follows from our model, let us come back to the
formula (\ref{k'k}). Let us assume for the moment that strings are
absent $(\sigma_{s0} =0)$ and, for example $k=0$. Then, from the
construction $4\sigma_{r0} + 3\sigma_{m0} - q_0 - 1$ and $k'=0$.
Next, we add strings $(\sigma_{s0}>0)$, leaving other astronomical
parameters $(\sigma_{r0}, \sigma_{m0},q_0)$ unchanged. Then, of
course $4\sigma_{r0} + 3\sigma_{m0} 2\sigma_{s0} - q_0 - 1>0$ and
consequently, the curvature index $k$ has to be equal to $1$.
However, $4\sigma_{r0} + 3\sigma_{m0} - q_0 - 1$ is still zero.
Hence, $k'=0$ and the dynamics of the universe, which depends on
the sign of $k'$, remains the same, although the curvature
changed. The same conclusion may also be deduced if $k=\pm 1$ from
the beginning. It seems that the reason for this property is that
the dynamics of the universe depends in principle (besides the $\Lambda$
term) on the sign of the expression $4\sigma_{r0} + 3\sigma_{m0} - q_0 -
1$, which does not include strings (see Eq. (\ref{H0R0})). For
completeness, it has to be stressed that adding an exotic fluid
other than the above one (for example, domain walls) changes the
dynamics because of the appearance of a qualitatively new term on
the right-hand side of Eq. (\ref{FRW2}).

Now we proceed to the presentation of the astrophysical formulas
for the universe with strings. A magnitude that enters most of the
expressions is a redshift $z$ of observed galaxies. We start with
the fundamental relation that establishes the connection between
the redshift and a radial coordinate $\chi$ of a galaxy. In the
most general case with cosmic strings, radiation, and a
cosmological term this relation is nontrivial and the Weierstrass
elliptic function ${\cal P}$ is used
\bea
\label{Pe}
{\cal P}(\chi) &=& \frac{k}{4\sigma_{r0} + 3\sigma_{m0} + 2\sigma_{s0} - q_0 -
1} \left( \frac{4\sigma_{r0} + 3\sigma_{m0} - q_0 - 1}{6} -
\frac{z+2}{2} \left[\sigma_{m0} + \sigma_{r0} (z+2) \right]
\right.\nonumber \\
&+& \left. \frac{1}{4z^2} \left\{ 1 + \left[2\sigma_{r0}z^2 (z+2)^2 +
\sigma_{m0}z^2(2z+3) + q_0 z(z+2) + (z+1)^2 \right]^{\frac{1}{2}}
\right\} \right)~.
\eea
The $\chi$ coordinate, by definition, is a difference between the
present value of the conformal time $\tau_0$ and a time $\tau$
corresponding to the moment of the emission of the light ray by
the galaxy with the redshift $z$:
\be
\chi = \tau_0 - \tau~.
\ee
If we compare the above formula with the appropriate one in the
model without strings (D\c{a}browski and Stelmach 1986b, 1987a), we
perceive that the difference is quite unremarkable. As a matter of
fact, it is no wonder, otherwise cosmic strings would have been
already discovered. Nevertheless, the distinction exists and
should come out in almost every astrophysical formula. Following
our last paper (D\c{a}browski and Stelmach 1987a), we specify some
expressions emphasizing particularly the universe with vanishing a
$\Lambda$ term and those that are string dominated.

\begin{center}
{\it a) The Luminosity Distance}
\end{center}

Denoting the right-hand side of Eq. (\ref{Pe}) by $f(z)$, the
formula for the luminosity distance of the observed galaxy is
\be
\label{D0}
D_0 = \frac{z+1}{H_0} \left( \frac{k}{4\sigma_{r0} + 3\sigma_{m0} + 2\sigma_{s0} - q_0 -
1}\right)^{\frac{1}{2}} S_k(\chi)~,
\ee
where
\be
\chi = {\cal P}^{-1} [f(z)]
\ee
and $S_k$ is defined
\bea
\label{rschi}
S_k(\chi) &=&\left\{
\begin{array}{l}
\label{Skchi}
\sin {\chi }\hspace{0.5cm}k>0, \\
\chi \hspace{0.5cm} k =0,  \\
\sinh {\chi }\hspace{0.5cm}k <0,
\end{array}
\right. \
\eea
In degenerate cases $(\lambda=0)$ the Weierstrass ${\cal P}$
function converts into an elementary one (Abramovitz and Stegun
1964):
\be
{\cal P}(\chi) = -\frac{k'}{12} + \frac{\mid k' \mid}{4 S_{k'}^2
\left(\frac{\mid k' \mid^{\frac{1}{2}}}{2} \chi \right)}~.
\ee
Then, the formula for $\chi$ reads
\bea
\label{chi}
\chi &=& \frac{1}{\mid k' \mid^{\frac{1}{2}}} S_{k'}^{-1} \left[ \frac{\mid k'
\mid^{\frac{1}{2}}}{z+1} \left(\frac{2\sigma_{r0} + 2\sigma_{m0} + 2\sigma_{s0} -
1}{k} \right)^{\frac{1}{2}}\right. \nonumber \\
& \times & \left. \frac{\sigma_{m0}z + (\sigma_{m0} + 2\sigma_{r0}
-1) ([2\sigma_{r0} z (z+2) + 2 \sigma_{m0} z + 1]^{\frac{1}{2}} -
1 )}{\sigma_{m0}^2 + 2\sigma_{r0}(2\sigma_{r0} + 2\sigma_{m0} -
1)} \right]~.
\eea
In models without strings $k'=k=0,\pm 1$, in consequence we have
\be
\frac{1}{\mid k' \mid^{\frac{1}{2}}} S_{k'}^{-1}(\mid k'
\mid^{\frac{1}{2}}\ldots ) = S_k^{-1}(\ldots )~.
\ee
Inserting Eq. (\ref{chi}) into (\ref{D0}) and remembering that
removing $\lambda$ allows elimination of $q_0$ (cf. Eq.
(\ref{lamob}), we get the formula for the luminosity distance in
a form that does not depend explicitly on curvature. In the
universe with strings, this is not the case. parameters $k'$ and $k$
generally do not coincide. Hence the composition
$S_k(\mid k' \mid^{-\frac{1}{2}}S_{k'}^{-1})$ does not yield the
identical function and the redshift-magnitude formula has a more
complicated form. All formulas significantly simplify only in
extreme cases. For example, choosing $\sigma_{r0}=0$, $\sigma_{m0}=\sigma_{s0} = 1/2$
and $k=1$, we have
\be
\chi = 2 \left( 1 - \frac{1}{\sqrt{z+1}} \right)~,
\ee
which coincides with the result known from the stringless
cosmology. Of course, this similarity is not accidental. The
reason is that in the case considered here the parameter $k'$,
which is some sense plays the role of the curvature index, is
equal to zero. Hence some cosmological effects following from the
existence of cosmic strings compensate for those typical for
closed geometry $(k=1)$. In spite of this, the formulas for the
luminosity distance are in both cases different (especially for
large $z$)
\bea
D_0 &=& \frac{z+1}{H_0} \sin{\left[2\left(1 -
\frac{1}{\sqrt{z+1}}\right)\right]}, \hspace{0.3cm}{\rm for} \hspace{0.3cm} k=1,
\sigma_{s0}=\frac{1}{2}~,\nonumber \\
D_0 &=& \frac{z+1}{H_0} 2\left(1 -
\frac{1}{\sqrt{z+1}}\right), \hspace{0.3cm}{\rm for} \hspace{0.3cm} k=0=
\sigma_{s0}=0~.
\eea
Another interesting case may be obtained when $\sigma_{s0}$ is
arbitrary but radiation and matter are negligible compared to the
density of strings. Then, the expression for $D_0$ is
\be
D_0 = \frac{z+1}{H_0} \left(\frac{k}{2\sigma_{s0}-1} \right)^{\frac{1}{2}}
S_k \left[\left(\frac{2\sigma_{s0}-1}{k} \right)^{\frac{1}{2}}
\ln{(z+1)}\right]~
\ee
and simplifies further by putting $k=0$
\be
D_0 = \frac{z+1}{H_0}\ln{(z+1)}~.
\ee
Note that adding the cosmological constant to the case $(\sigma_{r0}=\sigma_{m0}=0)$
does not lead beyond elementary functions (cf. Eqs.
(\ref{R1})-(\ref{R5})). In order to find $\chi$, we use the
formulas
\bea
\chi &=& \tau_0 - \tau = \int_{\tau}^{\tau_0} d\tau =
\int_{t}^{t_0} \frac{dt}{R(t)}~,\\
\frac{R_0}{R(t)} &=& z+1~,\\
R_0 &=& \frac{1}{H_0} \left(\frac{k}{2\sigma_{s0} - q_0
-1}\right)^{\frac{1}{2}}~,\\
k' &=& - (q_0 +1) \frac{k}{2\sigma_{s0} - q_0
-1}~,\\
\label{Lambdadeg}
\Lambda &=& - 3q_0H_0^2~,
\eea
where $R(t)$ is given by Eqs.(\ref{R1})-(\ref{R5}). performing
simple integrations, we get
\be
\label{chik'=0}
\chi = \sqrt{2\sigma_{s0}} z \hspace{0.3cm} {\rm for}
\hspace{0.3cm} k'=0~,
\ee
which implies that $k=1$, $q_0=-1$, $\Lambda=3H_0^2$~,
\be
\label{chik'>0}
\chi = \left(\frac{1+q_0-2\sigma_{s0}}{q_0+1}\right)^{\frac{1}{2}}
\left\{\arctan{\left(\frac{-1}{q_0+1}\right)^{\frac{1}{2}}} -
\arctan{\left[\frac{q_0}{(q_0+1)(z+1)^2} - 1
\right]^{\frac{1}{2}}}\right\} \hspace{0.3cm} {\rm for}
\hspace{0.3cm} k'>0~,
\ee
which implies that $k=1$, $q_0<-1$, $\Lambda>3H_0^2$, and
\be
\label{chik'<0}
\chi = \left(\frac{1+q_0-2\sigma_{s0}}{q_0+1}\right)^{\frac{1}{2}}
\ln{\frac{(z+1)(q_0+1)^{\frac{1}{2}} + [(q_0+1)(z+1)^2
-q_0]^{\frac{1}{2}}}{(q_0+1)^{\frac{1}{2}} +1}} \hspace{0.3cm} {\rm for}
\hspace{0.3cm} k'<0~,
\ee
which implies that $k=0,\pm 1$, $q_0>-1$, $\Lambda < 3H_0^2$. The
case where the $\Lambda$ term vanishes can be recovered only from
the last formula (\ref{chik'<0}) by putting $q_0=0$ (see Eq. (\ref
{Lambdadeg})).

\begin{center}
{\it b) Angular Diameter}
\end{center}

Writing an expression for the angular diameter $\theta$ of a
galaxy with the redshift $z$ in the universe with strings does not
lead to any difficulaties. We just use the known formula
\be
\label{angdiam}
\theta = \frac{d(z+1)}{D_0}~,
\ee
where $d$ is the linear size of the galaxy and $D_0$ its
luminosity distance given by Eq. (\ref{D0}). In this context, one
usually discusses minimal values of $\theta$, since it turns out
taht in the expanding universe the angular diameter of a galaxy is
not a decreasing function of its redshift. In the next section, we
shaw that $\theta_{min}$ depends distinctly on the density of
strings. Therefore, observation of $\theta_{min}$ may be a good
tool for testing a string-dominated universe.

\begin{center}
{\it c) Source Counts}
\end{center}

We calculate the number of sources with redshifts from the
interval $z$, $z+dz$. We use the same formula as in the standard
cosmology without strings,
\be
\label{dN}
dN = 4\pi n(z) S_k^2[\chi(z)]\frac{d\chi}{dz} dz~,
\ee
however, the $\chi$ coordinate is now given by Eq. (\ref{Pe}).
Such a general formula, although exact, is troublesome in
practical applications because its complicated form. Of course, in
some special cases the expression simplifies a little bit (cf.
Eqs. (\ref{chi}) and (\ref{chik'=0})-(\ref{chik'<0}), see also
Eq.(46) from our previous paper (D\c{a}browski and Stelmach 1987a).
However, notable amplification is obtained only by expanding the
appropriate expressions into series with respect to $z$. We find
that source counts are the strongest test for the cosmic strings
hypothesis. We shall discuss this point in detail in the next
section.

There exist some other cosmological formulas that may be found in
the models under considerations: particle and event horizon, the
age, or eventually maximum radius and period of oscillation of the
universe. In the quoted preprint (D\c{a}browski and Stelmach
1987b), we have derived these magnitudes and we discussed in
detail the influence of cosmic strings on them. In the present
paper, we do not want to pay too much attention to this. The task
of the first-rate significance is the verification of the
string-domination scenario on the basis of the strongest tests.

\section{Astrophysical Tests for the Cosmic Strings}

Writing down the solutions of the Friedmann equation in Sec. II we
realized that resulting cosmological models including strings
should be very similar to those without strings. We concluded that
the cosmic string network should be hardly detectable. To be more
precise, we consider now the formula for the luminosity distance $D_0$
to a galaxy with a redshift $z$. The complicated form of the
general formula (\ref{D0}) renders difficulties while fitting it
to observational data. Because of this and also because of the
small accuracy of the data it seems that using approximate
expressions for the relation is justified. We perform an expansion
of $D_0$ in the general case with radiation, matter, cosmological
constant, and strings into series with respect to the redshift.
The resulting expression is (see also Solheim 1966)
\be
D_0 = \frac{1}{H_0} \left\{z + \frac{1}{2} (1-q_0)z^2 +
\frac{1}{6} \left[3q_0(q_0 + 1) - 12 \sigma_{r0} - 6 \sigma_{m0} -
2 \sigma_{s0} \right]z^3 + \ldots \right\}~.
\ee
Note that $D_0$ is the luminosity distance, not a corrected one as
in the paper of Ellis and Maccallum (1970). The appearance of $\sigma_{s0}$
not before the third order of the expansion means that the
luminosity-distance relation is not a good test for the existence
of strings.

As a matter of fact, in astrophysical tests, instead of the
luminosity distance $D_0$ one rather measures the apparent
magnitude $m$ given by the formula (Lang 1974)
\be
m = 5\log{D_0} + M + {\rm const.}~,
\ee
where $M$ is the absolute magnitude. Expansion of the magnitude
(in the corrected form) into series (cf. Kristian and Sachs 1966,
Ellis and MacCallum 1970) gives
\bea
m &=& M - 5 \log_{10}{H_0} + 5 \log{z} \\
&+& (2.5 \log_{10}{e}) \left\{(1-q_0) z +
\left[\frac{1}{4} (3q_0 + 1)(q_0 - 1) - \frac{2\sigma_{s0}}{3} -
\frac{2\Lambda_{(4)}}{3H_0^2} \right] z^2 + \ldots \right\}.\nonumber
\eea
Here $\Lambda$ is given by Eq. (\ref{Lamob}). Note that the
radiation is excluded in the above formula by means of $\Lambda$.
here we notice that the strings appear in the second order. It
seems that the influence of strings is still weak. From the point
of view of the possibility of the detection of the cosmic strings,
this result is not promising.

Now we consider another test connected with the minimal angular
diameter of galaxies. The value of the redshift $z_{min}$ that
corresponds to $\theta_{min}$ has been calculated analytically for
pressureless Friedman models without the $\Lambda$ term and
strings (Edwards 1972, narlikar 1983), $z_{min} = \frac{5}{4}$. We
showed in our recent paper (D\c{a}browski and Stelmach 1987a) that
adding the radiation in a flat universe lowers this value:
\be
z_{min} = \frac{1}{4\sigma_{r0}}\left(2\sigma_{r0} - 1 +
\sqrt{6\sigma_{r0}} \right) \leq \frac{5}{4}~.
\ee
Calculation of $z_{min}$ in a general case, especially including
strings, is complicated because it leads to a non-elementary
equation,
\be
\label{SS}
\frac{S_k(\chi)}{S_k'(\chi)} = (z+1) \frac{d\chi}{dz}~.
\ee
It seems that an analytic solution exists only in one special case
when $k=\lambda=\sigma_{r0}=\sigma_{m0}=0$, i.e., in a
string-dominated universe. Then
\be
\chi = \ln{(z+1)}~
\ee
and Eq. (\ref{SS}) gives
\be
z_{min} = e - 1~.
\ee
In cosmological scale, the calculated value is remarkably greater
than the value $\frac{5}{4}$ obtained in a matter-dominated
universe. In order to see how $z_{min}$ changes if the universe
passes from the matter-dominated state to the string-dominated
one, a nonelementary equation has to be solved. For example, in
the case $k=\lambda=\sigma_{r0}=0$ (but $\sigma_{m0}\neq 0$ and $\sigma_{s0} \neq
0$), Eq. (\ref{SS}) takes the form ($\chi$ is given by Eq.
(\ref{chi}))
\bea
\sinh{\sqrt{\frac{2\sigma_{s0}}{z(1-2\sigma_{s0})+1}}} &=&
\frac{\sqrt{2\sigma_{s0}}}{\left(\frac{1}{2} -
\sigma_{s0}\right)^2}~\\
&\times&
\frac{z \left(\frac{1}{2} - \sigma_{s0} \right)
- \left(\sigma_{s0} + \frac{1}{2}
\right)\left(\sqrt{z(1-2\sigma_{s0}) + 1} - 1
\right)}{z+1}~.\nonumber
\eea
Its numerical solution with respect to $z$, for any value of
$\sigma_{s0} \in (0, \frac{1}{2})$ is presented in the form of a
diagram (\ref{fig1}).

The influence of the $\Lambda$ term on $z_{min}$ may also be
examined. We perform it for the string-dominated model. Then $\chi$
is given by Eq. (\ref{chik'<0}), and the equation to be solved is
\be
\ln{\frac{z+1 +[(z+1)^2 - q_0/(q_0 +1)]^{\frac{1}{2}}}{1+1/(q_0 +
1)}} = \frac{z+1}{[(z+1)^2 - q_0/(q_0+1)]^{\frac{1}{2}}}~.
\ee
In Fig. \ref{fig2} we illustrate the dependence of $z_{min}$ on
the deceleration parameter, which is proportional to the
cosmological constant in the case considered (cf. Eq.
(\ref{Lambdadeg})). Since the signs of the $\Lambda$ term and $q_0$
are opposite, we deduce from the diagram that the negative
cosmological constant, even when large, insignificantly lowers the
value of the redshift $z_{min}$, for which the engular size of a
galaxy is minimal, while the positive $\Lambda$ term dramatically
increases $z_{min}$. However, the influence of the small
cosmological constant may be ignored in any case.

At the end of our considerations, we propose another astrophysical
test, which distinguishes string- and matter-dominated models more
distinctly than the previous test. The test consists in counting
galaxies with a given redshift. Denoting this number by $N(z)$ and
coming back to the formula (\ref{dN}), we find that
\be
\label{N}
N(z) = 4\pi n(z) S_k^2[\chi(z)]\frac{d\chi}{dz} ,
\ee
We shall not examine this function in detail. As in the case of
the luminosity distance, we can expand it into series with respect
to $z$ (we put for simplicity $n(z)=$ const. $=n$.
\bea
\label{Nz}
N(z) &=& 4\pi n \left(\frac{4\sigma_{r0} + 3 \sigma_{m0} + 2\sigma_{s0} - q_0 -
1}{k}\right)^{\frac{3}{2}}  \\
&\times& \left\{z^2 - 2 (q_0 + 1)z^3 + \frac{1}{12}
\left[(q_0+1)(37q_0+31) - 96\sigma_{r0}
- 42\sigma_{m0} - 8 \sigma_{s0} \right]z^4
+ \ldots \right\} .\nonumber
\eea
In contradistinction to the previous case concerning $D_0$, now
strings appear in each order of the expansion, in a constant
factor. Coming back to the full formula (\ref{N}), we notice that
the function $N(z)$ possesses a maximum for some value $z_{max}$.
We are interested in only the dependence of $z_{max}$ on the
density of cosmic strings. In order to extract main features of
this dependence, we assume that $\lambda =\sigma_{r0}=k=0$. Then $\chi$
is given by Eq. (\ref{chi}). Differentiating $N(z)$ with respect
to $z$ and equating the result to zero, we find the following
formula for $z_{max}$:
\bea
\sinh{\frac{2\sqrt{2\sigma_{s0}[z(1-2\sigma_{s0})+1]}}
{(3z+1)(\frac{1}{2} - \sigma_{s0})}} &=&
\frac{\sqrt{2\sigma_{s0}}}{\left(\frac{1}{2} -
\sigma_{s0}\right)^2}~\\
&\times&
\frac{z \left(\frac{1}{2} - \sigma_{s0} \right)
- \left(\sigma_{s0} + \frac{1}{2}
\right)\left(\sqrt{z(1-2\sigma_{s0}) + 1} - 1
\right)}{z+1}~.\nonumber
\eea
In two limiting models (matter- and string-dominated universes),
the equation is solvable analytically, namely $z_{max}=16/9$ for $\sigma_{s0}=0$
and $z_{max}=e^2 - 1$ for $\sigma_{m0}=0$. Figure \ref{fig3} shows
the behaviour of $z_{max}$ as a function of $\sigma_{s0}$ in the
whole allowed interval $(0,\frac{1}{2})$. For comparison, we plot
also, in the same coordinate system, the dependence of $z_{min}$
on $\sigma_{s0}$. From the diagram, we see that although for small
density of strings the change of $z_{max}$ (compared to the
matter-dominated universe) is not remarkable, in the
string-dominated model the shift of $z_{max}$ is large. One may
show that the influence of the cosmological constant $z_{max}$ is
qualitatively the same as in the case of $\theta_{min}$.

\section{Conclusions}

The aim of the paper was to examine some observational
implications of the universe filled with a sufficiently dense
network of strings. The averaged energy density of strings was
assumed to scale as $\varrho_s \sim R^{-2}$. Hence, we assumed the
simplest possibility when the system of strings stretches
conformally by the expansion and may come to dominate the
universe. Besides cosmic strings, the universe contains
non-interacting pressureless dust, radiation, and a cosmological
constant. The mathematical formalism, which could describe such
general models, is based on Weierstrass functions. We have given
solutions of the Friedman equation and the principal astrophysical
formulas for observable quantities in the models. The first
conclusion from the results obtained was the confirmation of the
general conviction that the influence of the cosmic strings on the
evolution of the universe is relatively small in spite of the
sophisticated equation of state for the strings (pressure given by
strings is negative). On the basis of the derived formulas, we
proposed some tests, which may be used for the verification, if
not for the entire string scenario, at least for the
string-domination scenario (of course, only in the case $\varrho_s \sim
R^{-2}$. We showed that the luminosity distance and consequently
the apparent-magnitude relations are not sensitive tests if one
takes small accuracy of present observational data into account.
In the next test, we examined the influence of the cosmic strings
on the redshift $z_{min}$, for which the angular size of galaxies
takes a minimum value. We noticed that in the string-dominated
universe $z_{min}$ is shifted compared to the case without
strings, and the value of this shift is of order $\frac{1}{2}$. We
checked also that the small $\Lambda$ term (reasonable value for
our present universe) does not change the result too much. The
last test considered in the paper was connected with the source
counts. We proved that the number of sources corresponding to a
given redshift in a homogeneous, expanding universe possesses a
maximum depending strongly on the strings. Strictly speaking,
this dependence is not so distinct for a low energy density of
strings. Then $z_{max}$ is roughly equal to $2$. However, in
string-dominated universe, for example in the case when strings
form invisible dark matter giving $\Omega=1$, $z_{max}$
dramatically increases to $e^2 - 1$ for $\sigma_{s0}\to
\frac{1}{2}$, where $e=2.71828\ldots$. From Fig.\ref{fig3} it
follows that the string-dominated scenario (for example, where $90\%$
of the total energy is in strings and the rest is in matter)
should give $z_{max}=4$. One should also emphasize that $z_{max} \approx 2$
still allows existence of cosmic strings up to $40\%$ of the total
energy (in our considerations, we assumed that the number of
sources in a comoving volume is constant in time). In spite of the
strong dependence of $z_{max}$ on the density of strings
$\sigma_{s0}$, practical application of the presented test does
not seem to be possible up to now because of large values of
redshifts which should be taken into account $(z>2)$. First, the
problem of faint galaxies exists and there is a limit to how faint
a galaxy one can detect. The second problem is even more severe
and is connected with evolutionary effects of galaxies for large
$z$. The full formula for this test should involve the appropriate
corrections.

In conclusion, we would like to remind the reader that throughout
the paper we considered only the simplest system of cosmic strings
scaling as $\varrho_s \sim R^{-2}$. Discussion of observational
implications of other, maybe more realistic, string-domination
scenarios, where $\varrho_s \sim R^{-n}$ and $n>2$, will be
published in the future.


\appendix
\vspace{0.6cm}
\section{Astrophysical Formulas for the Cosmological Models with Domain
Walls} 

A redshift-magnitude relation for the universe which contains also
the domain walls (with $\varrho_w \propto R^{-1}$)
can be calculated following D\c{a}browski and Stelmach (1988)
\footnote{This appendix was not included in the original article, but for completness
of the discussion of the whole spectrum of cosmic fluids with the equations of state
from $p = -\varrho = -\Lambda$ to $p= \frac{1}{3}\varrho$ we
include the domain walls following D\c{a}browski, M.P., and Stelmach,
J. ((1988). In {\it Large Scale Structure in the Universe}, IAU Symposium
No. 130, edited by J. Audouze and A. Szalay (Reidel, Dordrecht)).}. 
Here we give the main result which generalizes our formula (\ref{Pe}) which is
a starting point for calculation all the observable quantities.
Firstly, one introduces a new constant $C_w$ which is related to
domain walls
\be
\label{Cw}
\varrho_w a  =  C_w \frac{3}{8\pi G}~,
\ee
and the dimensionless density parameter
\be
\sigma_{w0} = \frac{4\pi G \varrho_{w0}}{3H_0^2}~.
\ee
The Friedman equation written down in terms of dimensionless
variables {\ref{FRWT}) now takes the form
\begin{equation}
\label{FRWTw}
\left(\frac{dT}{d\tau}\right)^2=\alpha T^4+\frac{2}{3}T^3-k' T^2+\beta T+
\frac{\lambda}{3},
\end{equation}
where
\begin{equation}
\label{ceemetc}
\beta\equiv C_{w}\Lambda_c^{-1/2}.
\end{equation}
The solution of the Eq. (\ref{FRWTw})is now
\begin{equation}
\label{Ttauaw}
T(\tau) = \frac{3\sqrt{\alpha} {\cal P}'(\tau) - {\cal P}(\tau) - \frac{k'}{12}
- \frac{1}{48} \alpha \beta k'^2}{6 \alpha {\cal P}(\tau) - \frac{1}{6} -
k' \alpha}     ,
\end{equation}
where ${\cal P}$ is given by Eq. (\ref{Weierstrass}), i.e.,
\be
{\cal P}'(\tau) \equiv \frac{d{\cal P}}{d\tau} = \sqrt{4{\cal P}^3 -
g_{2}{\cal P} - g_{3}}   \nonumber   ,
\ee
and the invariants read as
\begin{eqnarray}
\label{g2w}
g_{2} & = & \frac{k'^2}{12} + \frac{\alpha\lambda}{3} - \frac{\beta}{6}  ,\\
\label{g3w}
g_{3} & = & 6^{-3} \left( k'^3 - 2\lambda - 3k'\beta \right) -
\frac{\alpha}{2} \left( \frac{k'\lambda}{9} + \frac{\beta^2}{8} \right)  .
\end{eqnarray}
In the model parameters (\ref{H0R0}), (\ref{k'k}), (\ref{Lamcob}),
(\ref{Lamob}), (\ref{lamob}) and (\ref{alphaob}) one has to
replace $(4\sigma_{r0} + 3\sigma_{m0} - q_0 - 1)$ by $(4\sigma_{r0} + 3\sigma_{m0} +
\sigma_{w0} - q_0 - 1)$ and $(4\sigma_{r0} + 3\sigma_{m0} + 2\sigma_{s0} - q_0 - 1)$
by $(4\sigma_{r0} + 3\sigma_{m0} + 2\sigma_{s0} + \sigma_{w0} - q_0 -
1)$. After that the formula (\ref{Pe}) generalizes to
\bea
\label{Pew}
&&{\cal P}(\chi) =
\frac{k}{4\sigma_{r0} + 3\sigma_{m0} + 2\sigma_{s0} + \sigma_{w0} - q_0 - 1} \times
\nonumber \\
& & \left( \frac{4\sigma_{r0} + 3\sigma_{m0} \sigma_{w0} - q_0 - 1}{6} -
\frac{z+2}{2} \left[\sigma_{m0} + \sigma_{r0} (z+2) \right]
\right. \\
&+& \left. \frac{1}{4z^2} \left\{ 1 + \left[2\sigma_{r0}z^2 (z+2)^2 +
\sigma_{m0}z^2(2z+3) - \sigma_{w0} z^2 + q_0 z(z+2) + (z+1)^2 \right]^{\frac{1}{2}}
\right\} \right)~.\nonumber
\eea
Inserting $\chi$ calculated as an inverse function of ${\cal P}$ from (\ref{Pew})
into (\ref{D0}), (\ref{angdiam}), and (\ref{dN}) allows to find
a redshift-magnitude relation, angular diameter, and the number of
galaxies with redshifts from $z$ to $z+dz$ in the universe which
is filled with all the cosmological fluids: radiation with $p_r = \frac{1}{3} \varrho_r$,
dust with $p_m = 0$, cosmic strings with $p_s = -\frac{1}{3}
\varrho_s$, domain walls with $p_w = \frac{2}{3} \varrho_w$, and
the cosmological constant with $p_{\Lambda} = - \varrho{\Lambda} = - \Lambda$.


\pagebreak

\begin{figure}[t]
\centering
\leavevmode\epsfysize=10cm \epsfbox{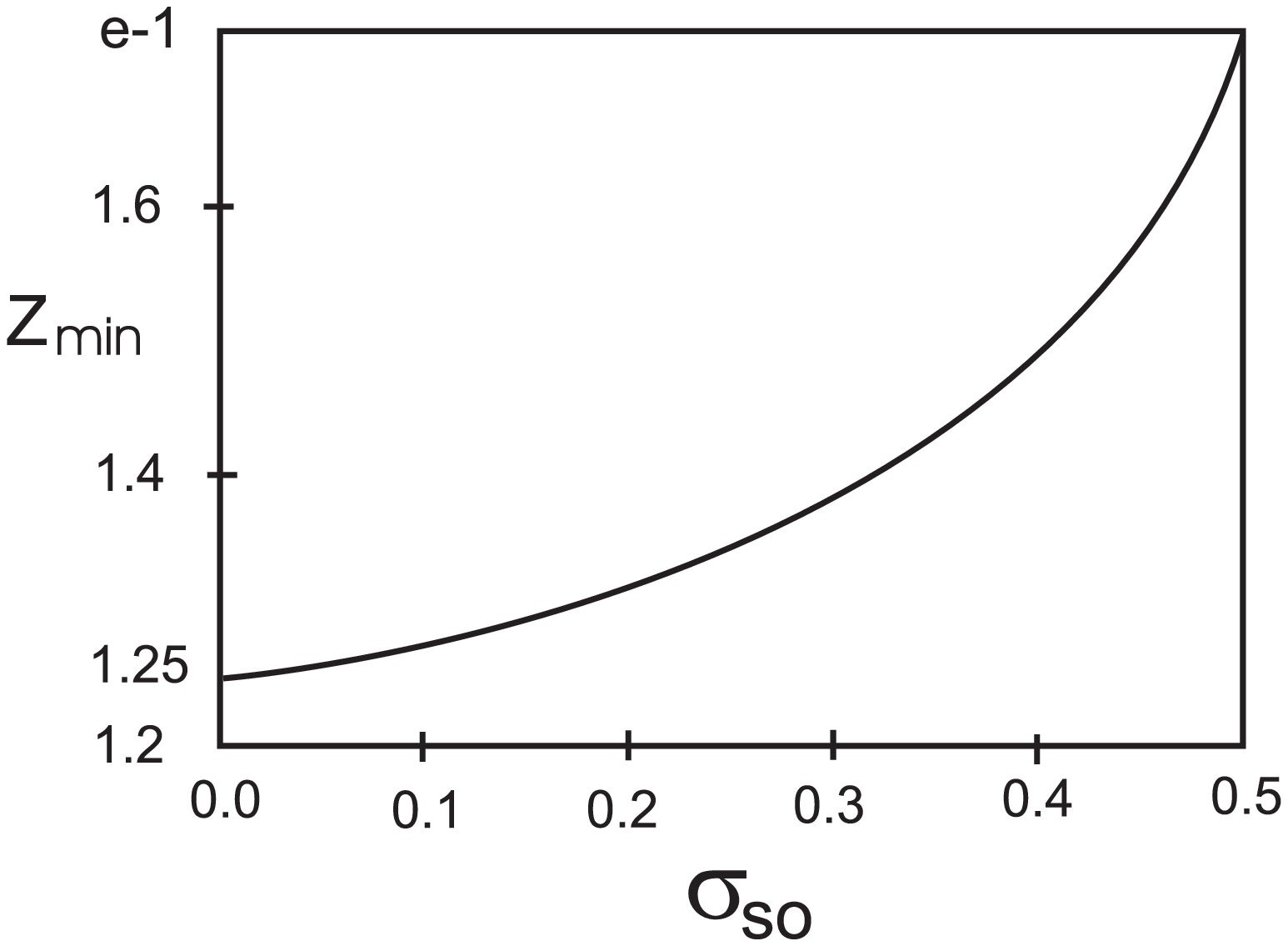}\\
\caption[]
{Dependence of the redshift, for which the regular size of
a galaxy takes a minimum value, on the energy density of strings.
$\lambda=k=\sigma_{s0}=0$, $\sigma_{m0}+\sigma_{s0}=\frac{1}{2}$.}
\label{fig1}
\end{figure}

\begin{figure}[t]
\centering
\leavevmode\epsfysize=10cm \epsfbox{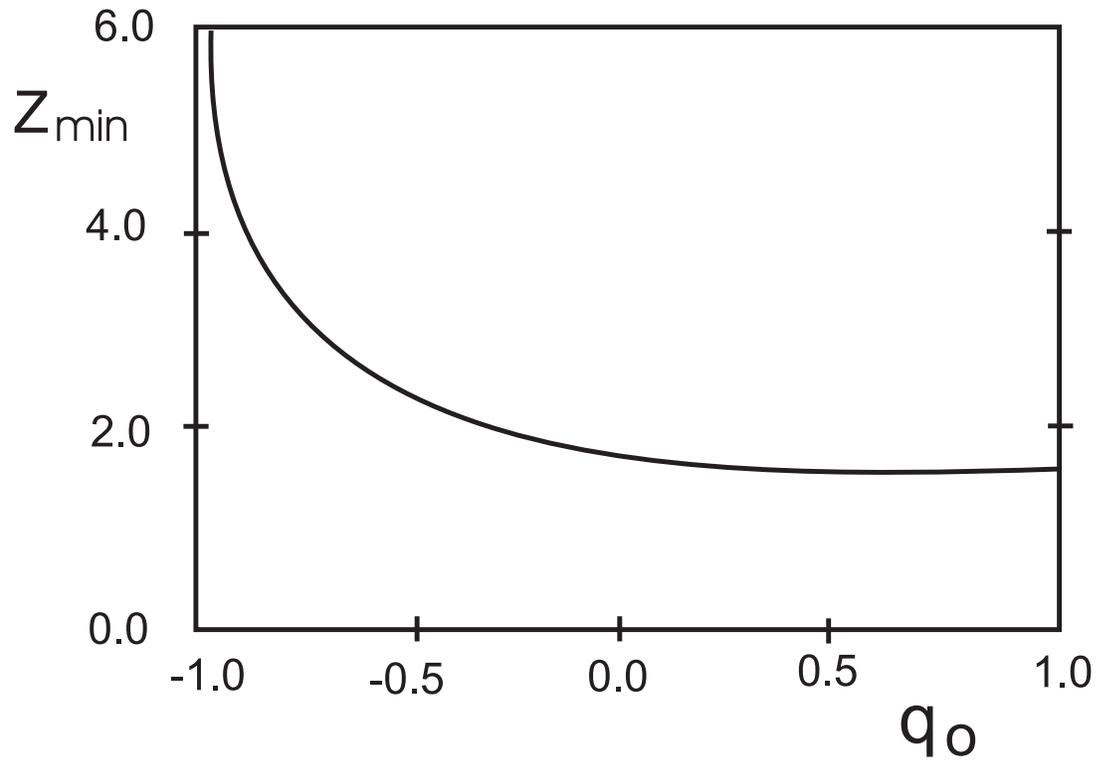}\\
\caption[]
{Influence of the cosmological constant on $z_{min}$.
$\sigma_{r0}=\sigma_{m0}=k=0$, $\sigma_{s0}=\frac{1}{2}$.}
\label{fig2}
\end{figure}

\begin{figure}[t]
\centering
\leavevmode\epsfysize=12cm \epsfbox{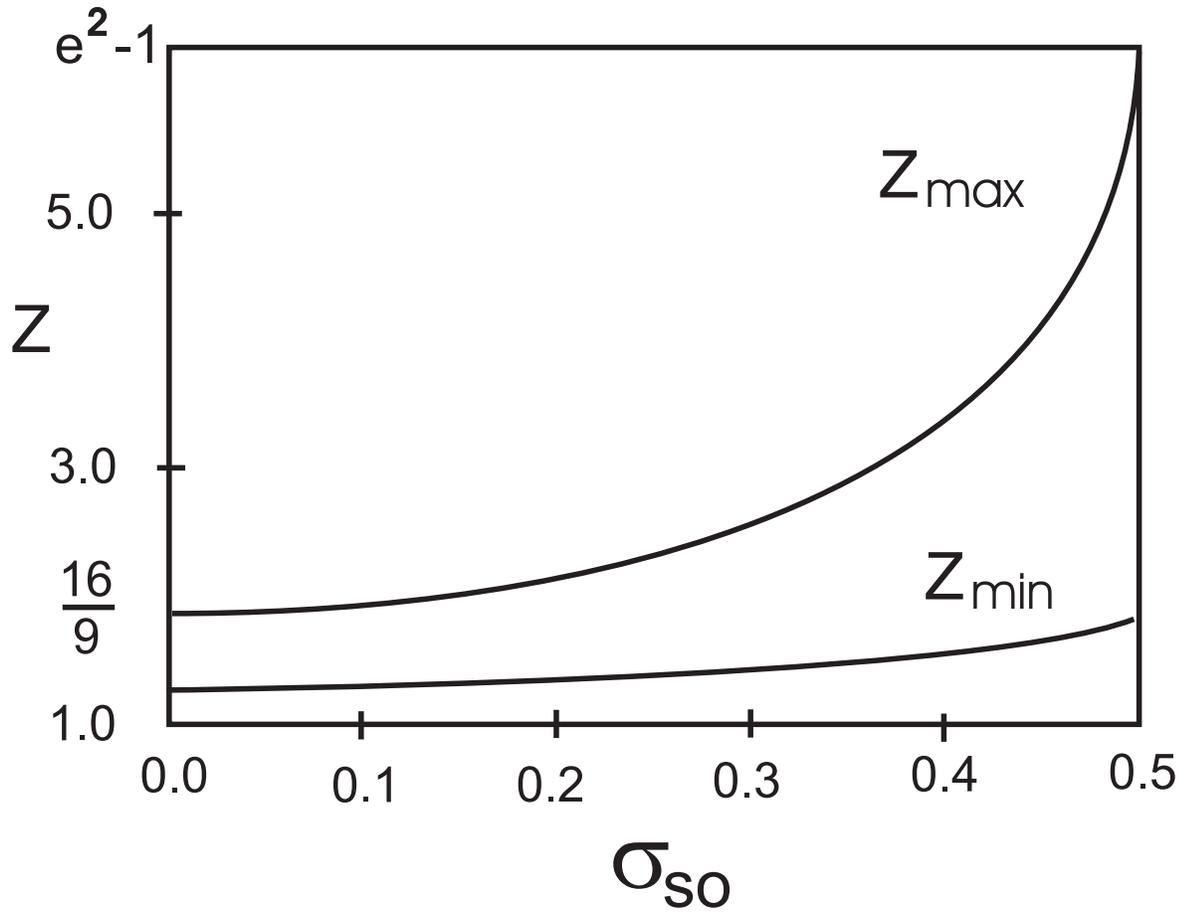}\\
\caption[]
{Dependence of the redshift, for which $N(z)$ takes a maximum
value, on the energy density of strings.
$\lambda=k=\sigma_{r0}=0$, $\sigma_{m0}+\sigma_{s0}=\frac{1}{2}$.}
\label{fig3}
\end{figure}

\vfill

\end{document}